\documentclass[]{spie}  %>>> use for US letter paper
\usepackage{amsmath,amsfonts,amssymb}
\usepackage{graphicx}
\usepackage[colorlinks=true, allcolors=blue]{hyperref}
\usepackage{float}

\title{The Space Coronagraph Optical Bench (SCoOB): X. Dark zone maintenance}

\author[a]{Saraswathi Kalyani Subramanian}
\author[a]{Kyle Van Gorkom}
\author[a]{Ramya M Anche}
\author[a,b]{William Melby}
\author[a]{Kian Milani}
\author[a,b]{Emory Jenkins}
\author[a]{Irina Stefan}
\author[a]{Adam Schilperoort}
\author[c]{Jaren N. Ashcraft}
\author[a,b]{Heejoo Choi}
\author[a,b]{Kevin Derby}
\author[a]{Olivier Durney}
\author[a,b]{Daewook Kim}
\author[a]{Kelsey Miller}
\author[a]{Ewan S. Douglas}
\affil[a]{Steward Observatory, University of Arizona, Tucson, AZ}
\affil[b]{James C. Wyant College of Optical Sciences, University of Arizona, Tucson, AZ}
\affil[c]{Department of Physics, University of California, Santa Barbara, CA}

\authorinfo{Further author information: (Send correspondence to S.K.S.)\\S.K.S.: E-mail: sksubramanian@arizona.edu}

\begin{document} 
\maketitle

\begin{abstract}
The Space Coronagraph Optical Bench (SCoOB) is a vacuum high contrast imaging testbed designed to demonstrate starlight suppression at optical wavelengths and obtain contrasts better than 10$^{-8}$ in a one-sided dark hole from 3 to 10 $\lambda/D$ using a vector vortex coronagraph (VVC) mask. Some of the recent efforts have been in testing dark zone maintenance (DZM) algorithms which are used to stabilize the contrast in the presence of wavefront drifts and allow for the long integration times required by exoplanet observations. In this work, we discuss the results from simulations with two DZM algorithms - linear dark field control (LDFC) and extended Kalman Filter-based (EKF) DZM. We also report preliminary results from the testbed with LDFC.

\end{abstract}

% Include a list of keywords after the abstract 

\section{INTRODUCTION}
\label{sec:intro}  % \label{} allows reference to this section
High contrast imaging (HCI) testbeds for space telescopes have been working toward demonstrating the high contrast required to image Earth-like planets around Sun-like stars\cite{2007Natur.446..771T}. This is achieved with coronagraphic optics typically made up of a focal plane mask (FPM) to suppress the on-axis starlight followed by a Lyot stop (LS) in the reimaged pupil plane to block the remaining diffracted starlight, in addition to using a DM to further suppress the residual light in a region of high contrast called the dark hole\footnote{The region of high contrast has also been called a dark field (DF) or dark zone (DZ). In this work, we use these three terms interchangeably to be consistent with the notation used for a specific algorithm in literature.}. 

A lot of effort in HCI testbeds so far has been directed toward improving the contrast; however, achieving high contrast is only the first step. Typical exoplanet observations can last tens of hours and due to thermal drifts or mechanical vibrations, the wavefront can drift and the contrast in the dark hole degrades. Therefore, we need to stabilize the DH contrast over extended periods of time. The Roman Space Telescope's Coronagraph Instrument (CGI) plans to handle this with an observation sequence as follows \cite{2020SPIE11443E..1UK}: point to a bright reference star and dig a dark hole, slew to the science target and observe for a few hours (during which the contrast degrades), then slew back to the reference star and re-dig the dark hole. While alternating between the sources restores the contrast, valuable science observation time is lost to slewing and re-digging the dark hole and every slew introduces further perturbations to the wavefront. Dark zone maintenance (DZM) algorithms are employed to stabilize the contrast on the science target, saving time by avoiding slewing as well as reducing the stability requirement of the telescope itself. This makes DZM crucial to the operation of future space-based coronagraphs and demonstrating these algorithms on HCI testbeds is important. 

The Space Coronagraph Optical Bench (SCoOB) \cite{maier_2020,ashcraft_2022} is a high-contrast imaging testbed operational at the University of Arizona since 2021. SCoOB uses a charge-6 VVC and a Kilo-C deformable mirror (DM) from Boston Micromachines (BMC) and achieved a contrast of $2.2\times10^{-9}$ in a $\ll 1 \%$ bandwidth (BW), $4\times10^{-9}$ in a 2\% BW, and $2.5\times10^{-8}$ in a 15\% BW\cite{vangorkom_2024} using iEFC (implicit Electric Field Conjugation)\cite{haffert_implicit_2023}. Previous studies have discussed the design \cite{ashcraft_2022}, development, end to end simulations\cite{anche_2025}, polarization aberration modeling and measurement\cite{anche2024space,ashcraft2024space}, demonstration of lyot-low order wavefront sensing (LLOWFS)\cite{milani2025space}, self-coherent camera (SCC)\cite{derby2025space} and adjoint EFC (aEFC)\cite{2025JATIS..11c9001M} on SCoOB. Recent upgrades to the hardware and the latest testbed configuration are presented in Van Gorkom et al.,: these proceedings\cite{kvg_2026}. In this work, we show the simulations and testbed demonstration of DZM algorithms on SCoOB.

We have considered two DZM algorithms -- linear dark field control (LDFC) \cite{2017JATIS...3d9002M, 2021A&A...646A.145M,2021A&A...653A..42B,2023arXiv230917449P} and the EKF-based (Extended Kalman Filter) DZM \cite{2019ApJ...873...95P, 2022JATIS...8c5001R, 10.1117/1.JATIS.12.4.041014}, hereafter referred to as just EKF in this work. In the next section, we discuss the working of the two algorithms followed by results from simulations for both. Finally, we show preliminary LDFC results from the testbed.

\section{Dark Zone Maintenance}
Wavefront sensing and control (WFS\&C) algorithms like EFC\cite{giveon_broadband_2007} and iEFC, that are used to dig the DH can improve the contrast over many orders of magnitude. However, these algorithms also perturb the DH, since the DM is probed to estimate the electric field, which can inject light into the DH and interfere with science observations. On the other hand, DZM algorithms are only required to stabilize the contrast with minimal perturbations to the DH to allow uninterrupted science observations.

\subsection{LDFC}
\label{subsec:ldfc}
LDFC uses the linear relationship between the intensity in the bright field (BF) and the wavefront perturbations to correct for the drifting wavefront error (WFE) and to stabilize the contrast in the DF. The intensity at the focal plane due to the initial field in the pupil plane ($E_0$) and changes in the field at the DM ($E_{DM}$) can be written as:

\begin{equation}
    I_t = |E_0|^2 + |E_{DM}|^2 + 2\langle E_0,E_{DM} \rangle. 
\end{equation}

The BF is dominated by the contribution from the initial electric field whereas the DF is dominated by the contribution from the changes by the DM. The intensity of the BF can therefore be approximated as:

\begin{equation}
    I_{BF} = |E_0|^2 + 2\langle E_0,E_{DM} \rangle,
\end{equation}

and the change in intensity at the BF with respect to the reference image (initial field) is:

\begin{equation}
    \Delta I_t = I_{BF} - I_{ref} \approx 2\langle E_0,E_{DM} \rangle.
\end{equation}

\begin{figure}[H]
    \centering
    \includegraphics[width=0.75\linewidth]{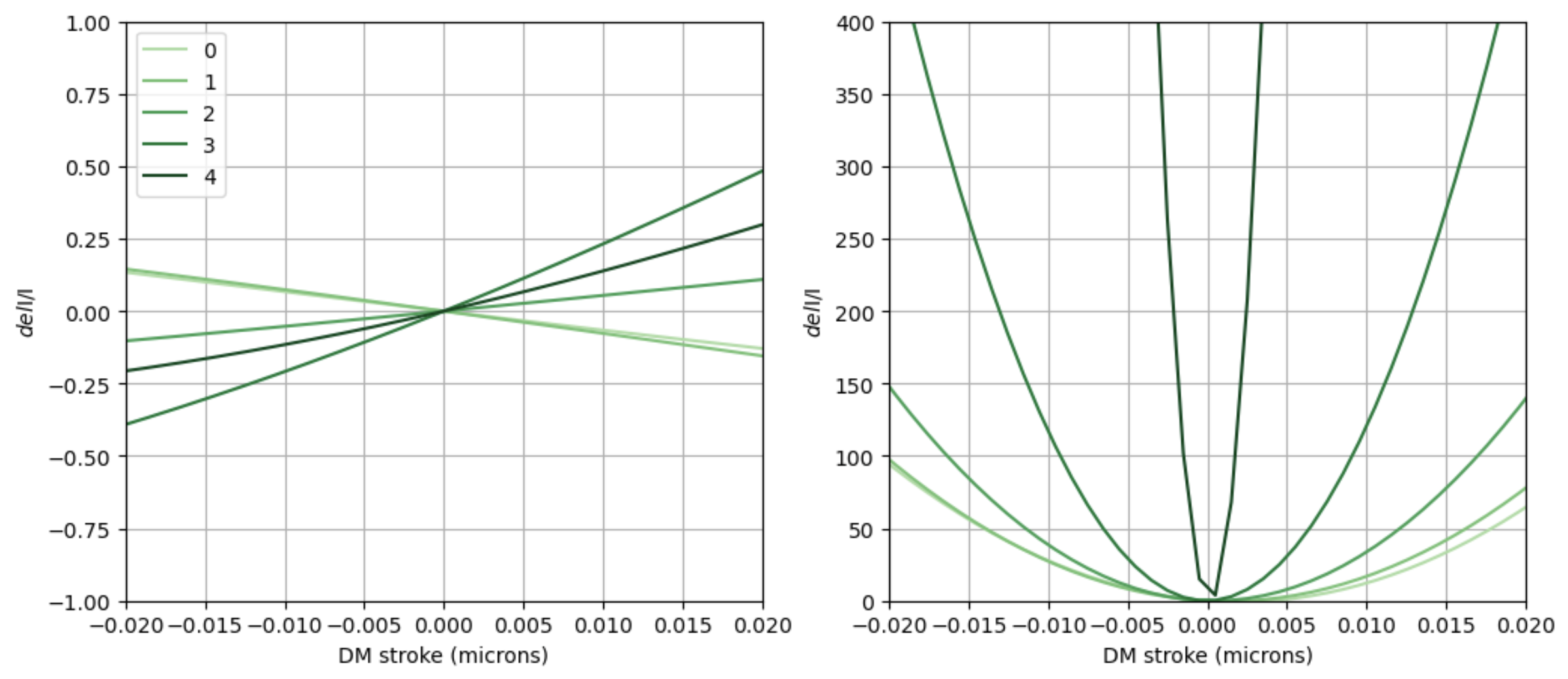}
    \caption{Difference in intensities between images recorded with only the DH command on the DM and with an added single actuator poked for a range of poke values for five pixels within the BF (left) and DF (right). These results are from the simulations described in Section~\ref{sec:sims}.}
    \label{fig:ldfc_bf_df_sim}
\end{figure}

Figure~\ref{fig:ldfc_bf_df_sim} show the difference in intensities between the unprobed and probed (single actuator poke) images at five pixels within the bright field and dark field, respectively. The linear relationship at BF pixels (Figure~\ref{fig:ldfc_bf_df_sim}: on the left) and the quadratic relationship at DF pixels (Figure~\ref{fig:ldfc_bf_df_sim}: on the right) can be observed.

\subsection*{Algorithm}
Since LDFC exploits the linear relationship between the wavefront and the intensity at the BF, it requires only one image per iteration. It is model-free and is typically employed with a two-stage calibration\cite{2021A&A...646A.145M,2021A&A...653A..42B}. The calibration steps are given below and the dimensions of the matrices used are given in Table~\ref{tab:ldfc_dimensions}.

   \begin{table}[ht]
\caption{
Dimensions of matrices used in LDFC. N$_\text{det}$ is the number of rows/columns of pixels in the detector, N$_\text{Hmodes}$ is the number of modes in the Hadamard basis, N$_\text{acts}$ is the number of actuators illuminated on the DM, and N$_\text{BF}$ is the number of pixels in the BF.
} 
\label{tab:ldfc_dimensions}
\begin{center}       
\begin{tabular}{|l|l|}
\hline
\rule[-1ex]{0pt}{3.5ex}  Matrix & Dimensions\\
\hline
\rule[-1ex]{0pt}{3.5ex}  I$_\text{ref}$ & N$_\text{det} \times$ N$_\text{det}$ \\
\rule[-1ex]{0pt}{3.5ex}  BF$_\text{mask}$ & N$_\text{det} \times$ N$_\text{det}$ \\
\rule[-1ex]{0pt}{3.5ex}  H$_\text{modes}$ & N$_\text{Hmodes}\times$N$_\text{Hmodes}$ \\
\rule[-1ex]{0pt}{3.5ex}   RM$_\text{HAD}$ &  N$_\text{Hmodes}\times$ N$_\text{det} \times$ N$_\text{det}$ \\
\rule[-1ex]{0pt}{3.5ex} IF$_{\text{cube,HAD}}$  & N$_\text{acts}\times$ N$_\text{det} \times$ N$_\text{det}$ \\
\rule[-1ex]{0pt}{3.5ex} IF$_{\text{cube,HAD,filt}}$  & N$_\text{acts}\times$ N$_\text{BF}$ \\
\rule[-1ex]{0pt}{3.5ex}  $\text{V*}_\text{HAD}$ & N$_\text{acts}\times$ N$_\text{acts}$\\
\rule[-1ex]{0pt}{3.5ex} RM$_{\text{EIG}}$  & N$_\text{acts}\times$ N$_\text{det} \times$ N$_\text{det}$ \\
\rule[-1ex]{0pt}{3.5ex}  IF$_{\text{cube,EIG}}$ & N$_\text{acts}\times$ N$_\text{det} \times$ N$_\text{det}$ \\
\rule[-1ex]{0pt}{3.5ex}  IF$_{\text{cube,EIG,filt}}$ &  N$_\text{acts}\times$ N$_\text{BF}$ \\
\rule[-1ex]{0pt}{3.5ex} CtrlMat & N$_\text{BF}\times$ N$_\text{acts}$ \\
\hline 
\end{tabular}
\end{center}
\end{table}

\begin{enumerate}
    \item Step 1: Set the PSF at the end of DH digging as the reference PSF (I$_{\text{ref}}$). LDFC will be used to drive the PSF back to this state. 
    
    \item Step 2: Define a BF pixel mask (BF$_\text{mask}$). The mask has N$_\text{BF}$ pixels set to ``True''. These can be set as the pixels where the contrast is greater than some threshold value (requires tuning based on instrument, flux, etc.). 
    
    \item Step 3: Apply the positive and negative of each mode of the chosen basis (we use Hadamard modes, H$_\text{modes}$) on the DM and the difference in intensities at each pixel of the detector forms a slice in the response matrix (RM$_{\text{HAD}}$).
    
    \item Step 4: The influence function cube (IF$_{\text{cube,HAD}}$) is then found as the product of RM$_\text{HAD}$ and the truncated inverse of H$_\text{modes}$ (truncated to the number of illuminated actuators). This inverts RM$_\text{HAD}$ into the response of the wavefront sensor (WFS) to the DM actuator space.
    
    \item Step 5: Derive the second basis set, called the ``DM eigenmodes'' ($\text{V*}_\text{HAD}$) by doing a singular value decomposition (SVD) of the transpose of the filtered IF$_{\text{cube,HAD}}$ (accounting only for the pixels in the BF mask). 
    \[  \text{IF}_{\text{cube,HAD,filt}}^{T} = \text{U}_\text{HAD}\text{S}_\text{HAD}\text{V*}_\text{HAD} . \]

    \item Step 6: Repeat Step 3 but with the DM eigenmodes as the new basis, i.e., the difference in intensities in all the pixels in the detector to the positive and negative of each DM eigenmode forms the slices of the response matrix of DM eigenmodes (RM$_{\text{EIG}}$).

    \item Step 7: Find the DM eigenmode influence function cube (IF$_{\text{cube,EIG}}$) from RM$_{\text{EIG}}$. This is the product of RM$_{\text{EIG}}$ and the inverse of the DM eigenmodes.

    \item Step 8: Find the LDFC control matrix (CtrlMat) by inverting the filtered IF$_{\text{cube,EIG,filt}}$ (taking only the BF pixels into consideration).  
    
\end{enumerate}

With the CtrlMat, the LDFC closed loop operation follows that of a conventional AO control loop. In each iteration, an image is recorded at the science camera. The intensities of the pixels within the BF$_\text{mask}$ are multiplied with the CtrlMat to determine the commands to be sent to the DM actuators. Thus, each loop only requires one image per iteration, and since we use intensities and not the electric field, no additional probes on the DM are required and the DF is left undisturbed by the DZM algorithm itself. 

\subsection{EKF}
The Extended Kalman Filter (EKF) is a non-linear estimator that solves for the maximum likelihood estimate and is commonly used by many control systems. At each iteration, it combines an estimate of the state (and by extension a prediction of the measurement) with the actual measurement to refine the estimate of the state.  

The state of our system is the real and imaginary parts of the drifting electric field ($x^k$) at each pixel within the dark hole. It can be controlled by commanding the DM ($u^k$) and the measurements are the images from the science camera ($z^k$). These values are initialized as follows: we assume that the electric field has not drifted and that the state/open loop electric field is the field at the end of DH digging. Therefore, no control is given to the DM. EKF works by iterating over the five equations given below to obtain an estimate of the electric field.

\begin{enumerate}
    \item Step 1: Using only the system model, extrapolate the state and covariance matrix ($P^k$) estimates. Since our system is static and the drift in the WF is caused by external factors, the state transition matrix ($F^k$) is an identity matrix.

    \[ \hat{x}^{k|k-1} = F^k\hat{x}^{k-1} \]
    \[ P^{k|k-1} = F^kP^{k-1}(F^k)^T + Q^k \]

    Here, $Q^k$ is the process noise, which is determined by the drifts in our system. 

    \item Step 2: The Kalman gain is estimated as:
    \[ K^k = P^{k|k-1}(H^k)^T(H^kP^{k|k-1}(H^k)^T + R^k)^{-1}, \]

    where $H^k$ is the observation matrix linearized at each iteration. It is the transformation of the state estimate to measurements ($H^k = \frac{\partial z^k}{\partial x^k} = \left[ 2\mathfrak{R}\left\{ x^k \right\} \quad 2\mathfrak{I}\left\{ x^k \right\} \right] $). $R^k$ is the measurement noise (or the sensor noise). 

    \item Step 3: Now, the state and covariance matrices are updated using the previous state estimate (from Step 1) as well as the measurement ($z^k$):
    \[ \hat{x}^k = \hat{x}^{k|k-1} + K^k \big[z^k - h(\hat{x}^{k|k-1}, u^k) \big] \]
    \[ P^k = P^{k|k-1} - K^kH^kP^{k|k-1}, \]
    where $h(\hat{x}^{k|k-1}, u^k)$ is the estimated measurement from the predicted state. 
\end{enumerate}

Like LDFC, each iteration of EKF utilizes only one ``measurement'' or one image and it uses the above equations to get an estimate of the electric field. While there are no probes, typically a small dither is added to the DM to increase the phase diversity of the measurements. The EFC Jacobian can then be used to convert the field estimate into DM commands to control the drifting field. 

   \begin{table}[H]
\caption{
Dimensions of the matrices used by EKF. N$_\text{DZ}$ is the number of pixels within the DH and 
N$_\text{acts}$ is the number of actuators illuminated on the DM.} 
\label{tab:ekf_dimensions}
\begin{center}       
\begin{tabular}{|l|l|}
\hline
\rule[-1ex]{0pt}{3.5ex}  Matrix & Dimensions\\
\hline
\rule[-1ex]{0pt}{3.5ex} $\hat{x}$ & 2N$_\text{DZ}$ \\
\rule[-1ex]{0pt}{3.5ex} $u$ & N$_\text{acts}$ \\
\rule[-1ex]{0pt}{3.5ex} $z$ & N$_\text{DZ}$ \\
\rule[-1ex]{0pt}{3.5ex} $P$ & 2N$_\text{DZ}\times$2N$_\text{DZ}$ \\
\rule[-1ex]{0pt}{3.5ex} $Q$ & 2N$_\text{DZ}\times$2N$_\text{DZ}$ \\
\rule[-1ex]{0pt}{3.5ex} $H$ & N$_\text{DZ}\times$2N$_\text{DZ}$ \\
\rule[-1ex]{0pt}{3.5ex} $R$ & N$_\text{DZ}\times$N$_\text{DZ}$ \\
\hline 
\end{tabular}
\end{center}
\end{table}

\section{Simulations}
\label{sec:sims}
 Using a model of SCoOB in \texttt{POPPY} (Physical Optics Propagation in PYthon) \cite{poppy}, we performed simulations of both LDFC and EKF algorithms.  The simulations were done at monochromatic wavelengths and use a Fresnel model of SCoOB given as a .toml file (described in Van Gorkom et al.,: these proceedings\cite{kvg_2026}) that accounts for the surface error measurements. We have not considered detector noise for these simulations and it will be accounted for in future work. 

\subsection{LDFC}
Following the steps described in Section~\ref{subsec:ldfc}, we perform the two-stage calibration of LDFC and obtain a CtrlMat. The DM eigenmodes ($V^*_\text{HAD}$) from simulations are shown in Appendix~\ref{sec:app_dm_eig} (Figure~\ref{fig:ldfc_dm_eig_modes}).

\begin{figure}[H]
    \centering
    \includegraphics[width=0.95\linewidth]{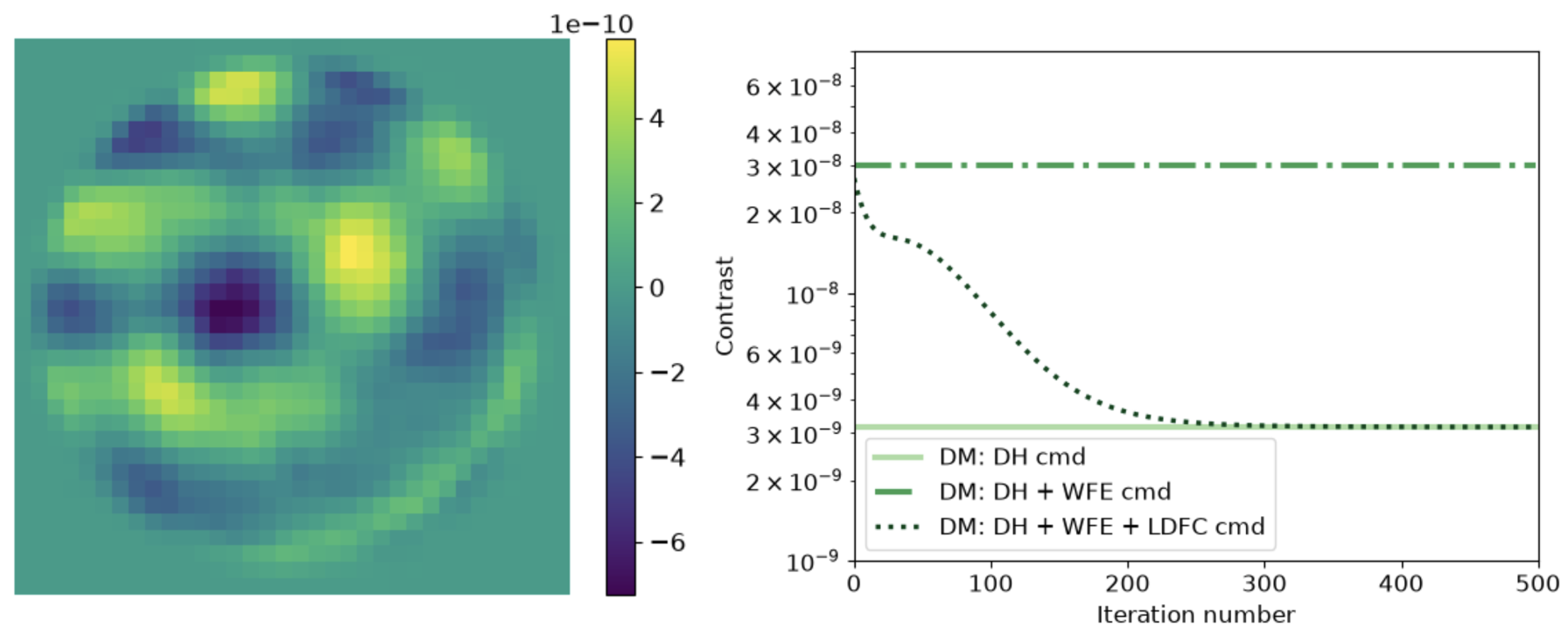}
    \caption{Left: A single DM eigenmode is introduced as the static WFE. Right: Final DH contrast at the end of iEFC (solid line), contrast curve during LDFC operation as it closes the loop (dotted line) and contrast with an uncorrected static WFE (dashed line).}
    \label{fig:ldfc_sim_static_dm_eig_mode}
\end{figure}

We first tested the algorithm by introducing a static WFE made of a single DM eigenmode. Figure~\ref{fig:ldfc_sim_static_dm_eig_mode} (left) is the static WFE introduced, which causes the contrast to jump to about 3$\times$10$^{-8}$ (dashed curve on the right). When LDFC is used, the contrast curve converges (dotted line) to the value that we had at the end of iEFC (solid line).

\begin{figure}[H]
    \centering
    \includegraphics[width=0.95\linewidth]{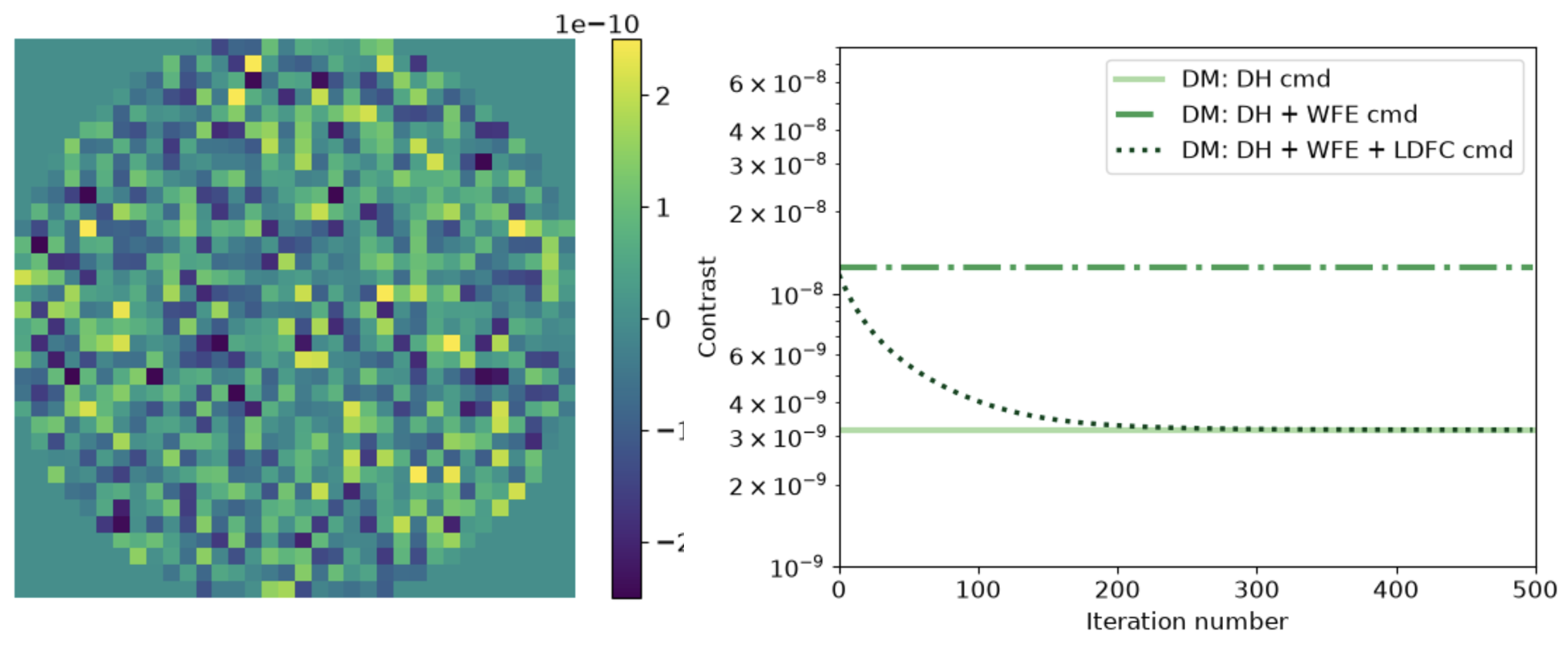}
    \caption{Similar to Figure~\ref{fig:ldfc_sim_static_dm_eig_mode} but with a static WFE that is a random walk of DM actuators.}
    \label{fig:ldfc_sim_static_rnd_wlk}
\end{figure}

Next, we tested with a random walk of DM actuators. The random walk is written as:

\begin{equation}
\label{eq:rnd_wlk}
     u^{k+1}_\text{drift} = u^{k}_\text{drift} + N(0,\sigma^2_\text{drift}I),
\end{equation}

where $\sigma_\text{drift}$ is the standard deviation of the normal distribution ($N$) from which values are drawn to update the random walk of each actuator. $I$ is the identity matrix with dimensions matching the number of actuators. Assuming a drift of 0.1~nm, we update the random walk equation for one iteration. We then freeze the positions of the actuators and consider it to be a static WFE, and the results from that are shown in Figure~\ref{fig:ldfc_sim_static_rnd_wlk}.

\begin{figure}[H]
    \centering
    \includegraphics[width=0.95\linewidth]{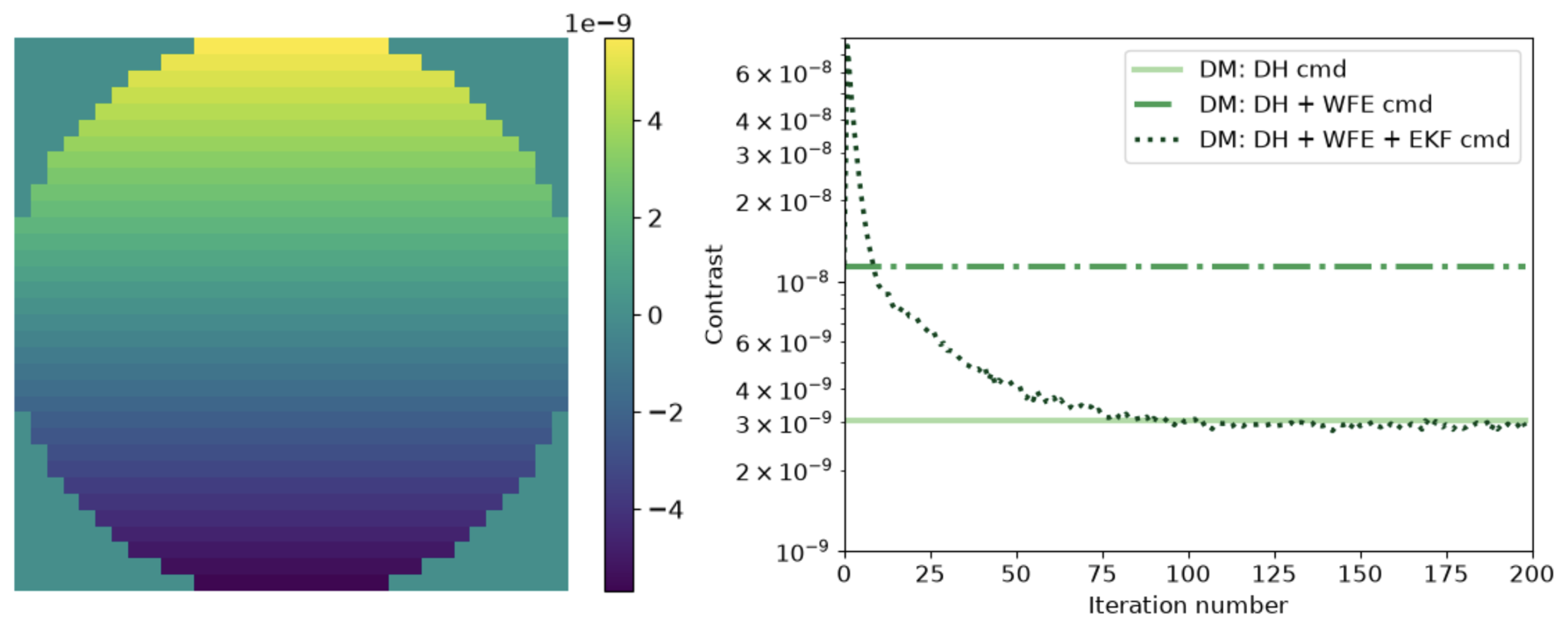}
    \caption{Left: Static Zernike WFE. Right: Final iEFC DH contrast (solid line), contrast with static WFE (dashed line), and contrast during EKF control loop (dotted line).}
    \label{fig:ekf_sim_static_zernike}
\end{figure}

\subsection{EKF}
Similarly, we first tested the EKF algorithm with a static Zernike mode as the WFE. The WFE is shown in Figure~\ref{fig:ekf_sim_static_zernike} (left). On the right of the figure, the ideal DH contrast without any drift (solid line), contrast in the presence of a static WFE (dashed line) and the curve as EKF stabilizes the contrast (dotted line) are shown. The EKF algorithm takes a few iterations to converge on a solution. This can be seen by the worsening contrast in the first few iterations. In practice, the gain of the control loop is set to zero for the first 10-20 iterations for this reason. 

\begin{figure}[H]
    \centering
    \includegraphics[width=0.95\linewidth]{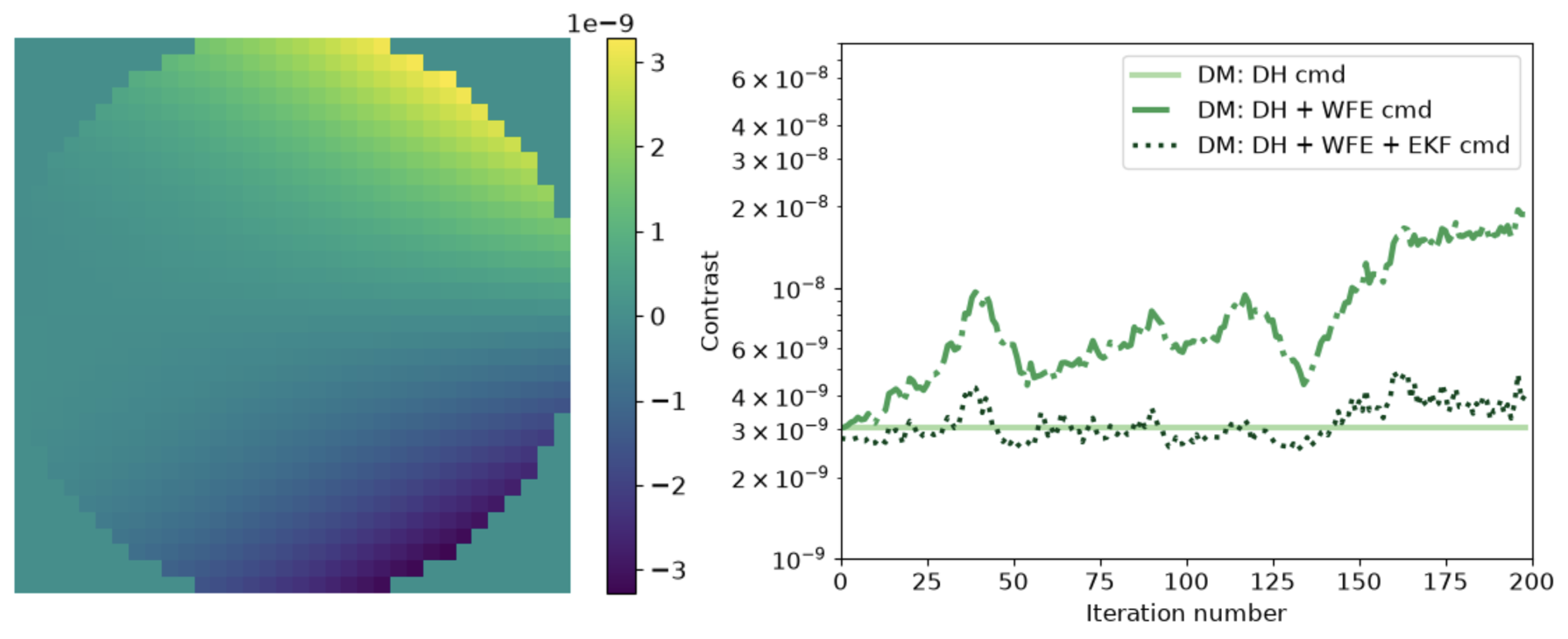}
    \caption{Left: One frame from the dynamic WFE made up of a sum of lower order Zernikes. Right: Final iEFC DH contrast (solid line), contrast with the accumulating WFE (dashed line), and contrast during EKF control loop (dotted line).}
    \label{fig:ekf_sim_dynamic_zernike}
\end{figure}

\begin{figure}[H]
    \centering
    \includegraphics[width=0.95\linewidth]{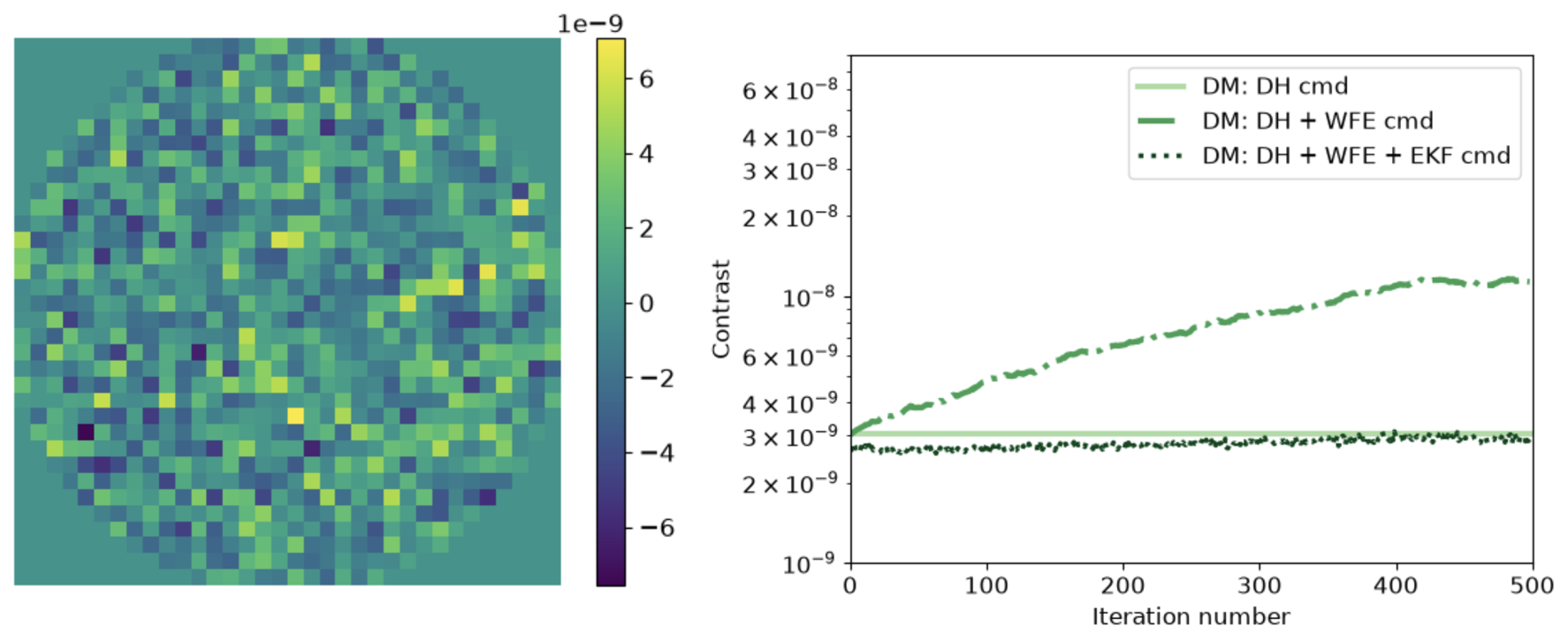}
    \caption{Similar to Figure~\ref{fig:ekf_sim_dynamic_zernike} but with a random walk of DM actuators as the WFE.}
    \label{fig:ekf_sim_rnd_wlk}
\end{figure}

Following this, we considered a scenario where the RMS of the WFE slowly accumulates over time. We simulated this by multiplying our WFE (in this case, a sum of a few lower order Zernikes of varying amplitudes) by a random amplitude factor that grew in each iteration. Here, as seen on the dashed curve in the right of Figure~\ref{fig:ekf_sim_dynamic_zernike}, the contrast slowly degrades over time. As the dotted curve shows, EKF is able to maintain it at the level of the final iEFC contrast.

We repeated the same with a random walk described in Equation~\ref{eq:rnd_wlk}, and the results are shown in Figure~\ref{fig:ekf_sim_rnd_wlk}.

\section{Preliminary testbed results}

\begin{figure}[H]
    \centering
     \includegraphics[width=0.85\linewidth]{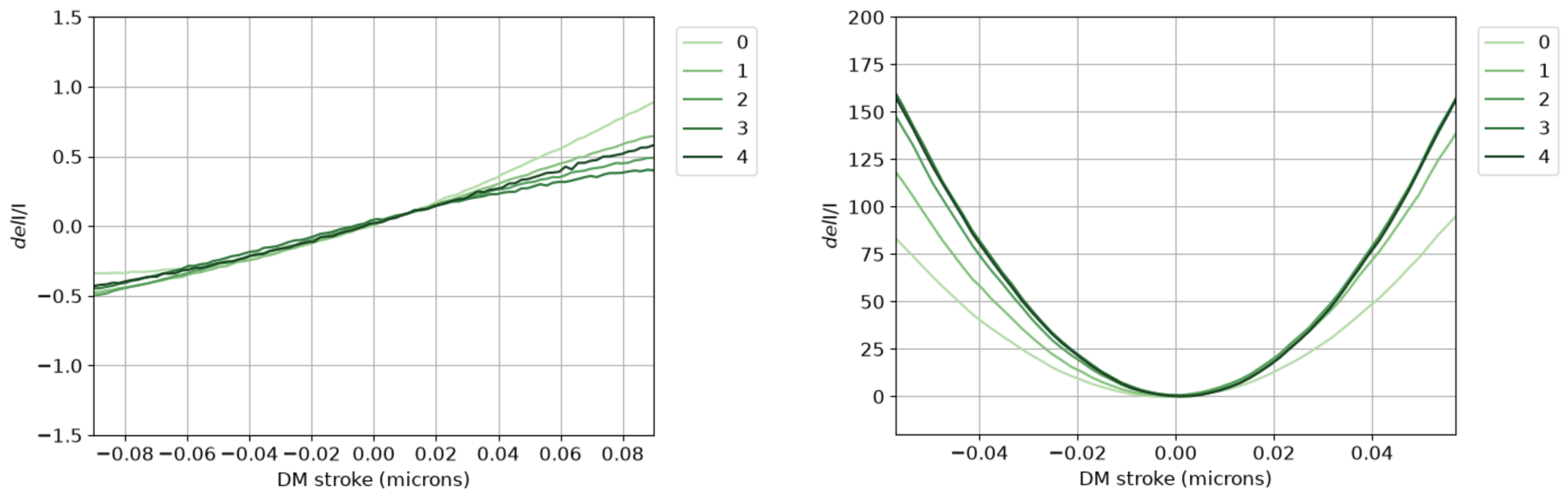}
    \caption{BF and DF response curves similar to Figure~\ref{fig:ldfc_bf_df_sim} recorded on the bench.}
    \label{fig:LDFC_BF_DF_response_scoob}
\end{figure}

We have performed an initial demonstration of LDFC on SCoOB in-air in the thermal vacuum chamber (TVAC). The components used during the experiments described below are listed at \texttt{scoob\_state} github respository at \href{https://github.com/uasal/scoob_state/commit/af75b159299008c0009ab2b7bdc3b972a18dd28a#diff-91f8faf7714ad1bfb2ac6cb4a13043fd0d249b8e5c6b6338b81f86ccf399ebfaR217}{scoob\_state@af75b15}.
The change in response at the BF and DF pixels to a single actuator poke as recorded on the testbed is shown in Figure~\ref{fig:LDFC_BF_DF_response_scoob}.

\begin{figure}[H]
    \centering
    \includegraphics[width=0.95\linewidth]{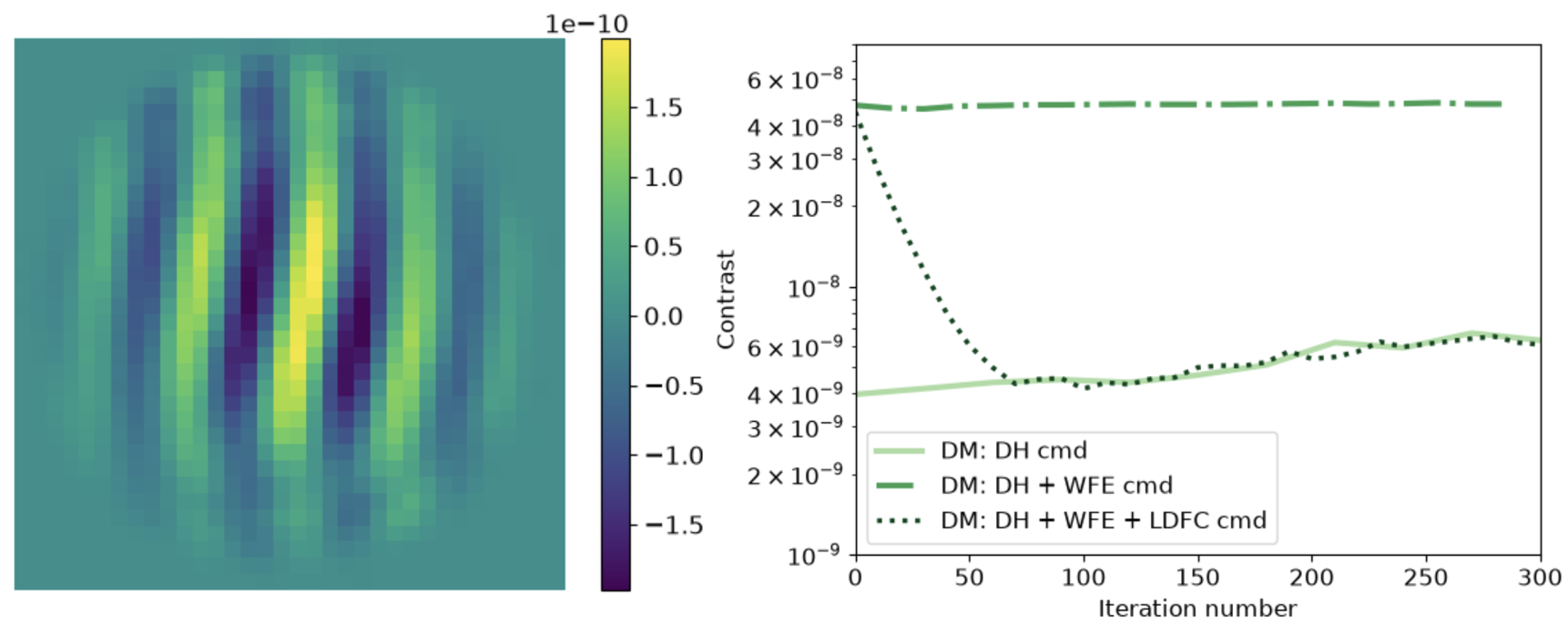}
    \caption{Left: Static eigen mode that is added as the WFE on the DM. Right: The contrast with only the final iEFC DM command (solid line), contrast with WFE added on the DM (dashed line), contrast as LDFC corrects for the added WFE (dotted line).}
    \label{fig:ldfc_static_eig_mode_scoob}
\end{figure}

As a first test of the algorithm, we injected a static WFE made of a single eigenmode after using iEFC to dig a DH (monochromatic contrast at 630~nm: 3$\times$10$^{-9}$). This is similar to one of the test cases that we simulated. The static WFE we added is shown on the left of Figure~\ref{fig:ldfc_static_eig_mode_scoob}. Similar to the other figures, the plot on the right shows the contrast with just the final iEFC DM command (solid line), contrast with WFE added on the DM (dashed line), and the contrast as LDFC corrects for the added WFE (dotted line). The contrast curves with only the DH command and the one with WFE added were recorded only every 30 and 15 iterations, respectively. It can be seen that there is some drift in the testbed which LDFC is unable to correct for. The drift appeared as a lower order WFE (the airy rings could be seen leaking through into the DH with time) which we found LDFC to be insensitive to, even in simulations. However, LDFC was able to correct for the injected WFE.

\begin{figure}[H]
    \centering
    \includegraphics[width=0.95\linewidth]{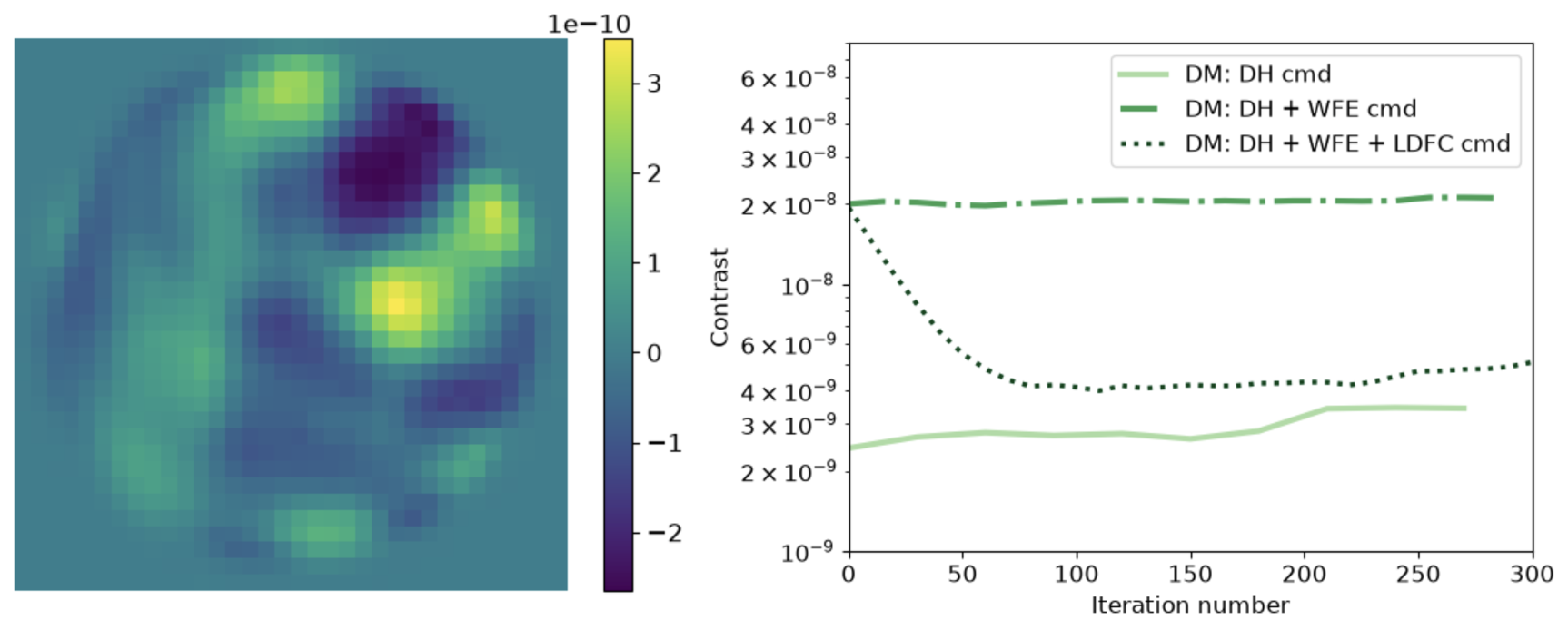}
    \caption{Similar to Figure~\ref{fig:ldfc_static_eig_mode_scoob} but with the WFE being a sum of DM eigenmodes.}
    \label{fig:ldfc_static_sum_eig_modes_scoob}
\end{figure}

Similarly, we tested with a static WFE that was made of a sum of few low order DM eigenmodes. The results are shown in Figure~\ref{fig:ldfc_static_sum_eig_modes_scoob}.

\section{Conclusions}
We have tested two different DZM algorithms using simulations and have also recorded some preliminary results of DZM on the testbed. We have found that while LDFC is able to correct for the injected drift (in the form of DM eigenmodes), it was unable to correct for the drift of the testbed itself. Simulations also suggest that LDFC might be less sensitive to low-order Zernikes when used with a VVC. We are exploring other forms of calibration for LDFC to improve the sensitivity to lower-order modes. Our next steps are to test the EKF-based DZM on the bench and to run LLOWFS and DZM simultaneously to stabilize the contrast in our DH.

\appendix 
\section{DM Eigen modes}
\label{sec:app_dm_eig}

\begin{figure}[H]
    \centering
    \includegraphics[width=0.85\linewidth]{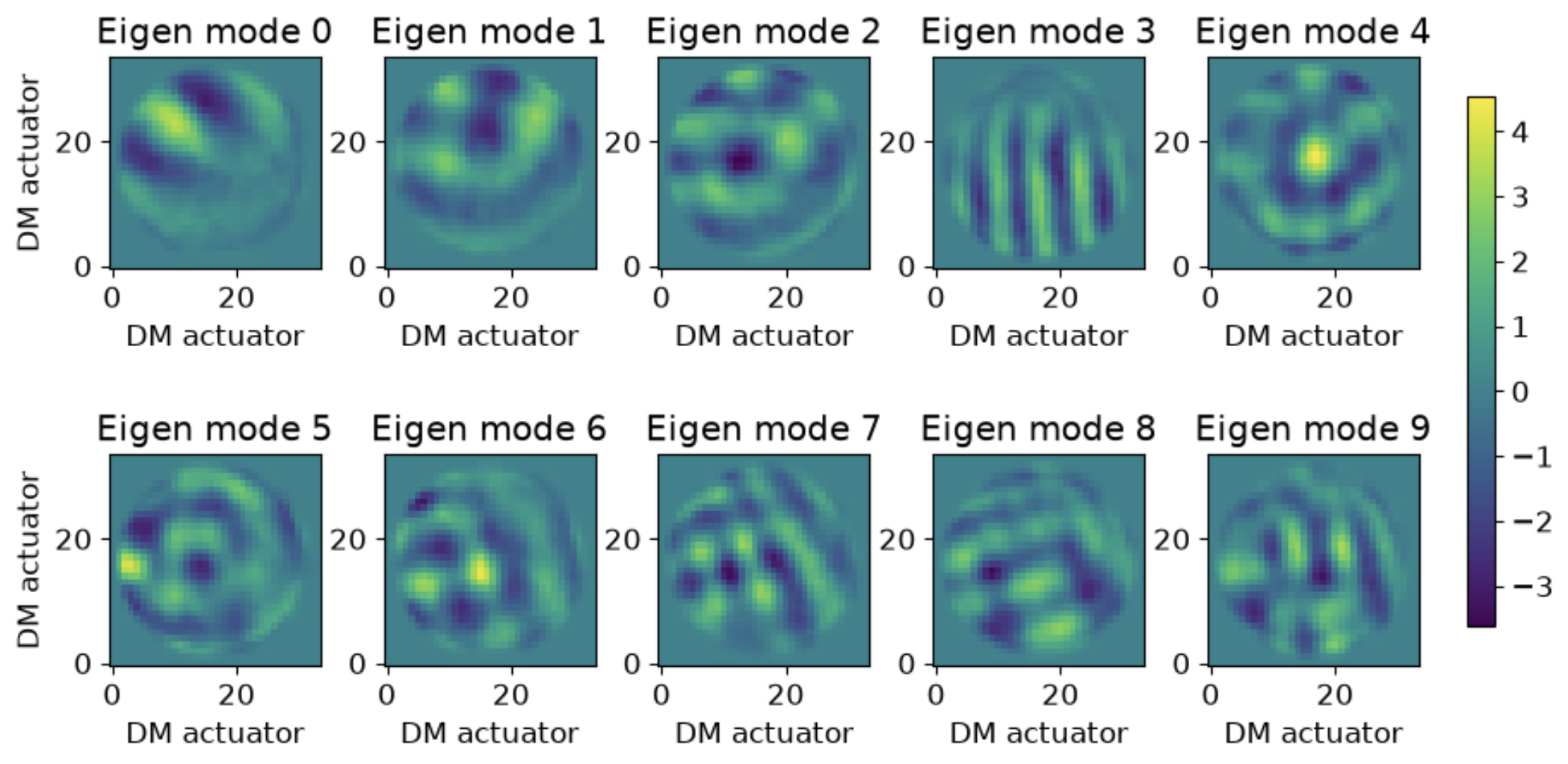}
    \caption{The first ten DM eigenmodes from simulations. The second calibration for LDFC is done with these modes as the basis.}
    \label{fig:ldfc_dm_eig_modes}
\end{figure}

\section*{Acknowledgments}
Portions of this research were supported by funding from the Technology Research Initiative Fund (TRIF) of the Arizona Board of Regents and from Schmidt Sciences. J.N.A was supported by NASA through the NASA Hubble Fellowship grant \#HST-HF2-51547.001-A awarded by the Space Telescope Science Institute, which is operated by the Association of Universities for Research in Astronomy.

% References
\bibliography{report} % bibliography data in report.bib

@inproceedings{giveon_broadband_2007,
	location = {San Diego, {CA}},
	title = {Broadband wavefront correction algorithm for high-contrast imaging systems},
	url = {http://proceedings.spiedigitallibrary.org/proceeding.aspx?doi=10.1117/12.733122},
	doi = {10.1117/12.733122},
	eventtitle = {Optical Engineering + Applications},
	pages = {66910A},
	author = {Give'on, Amir and Kern, Brian and Shaklan, Stuart and Moody, Dwight C. and Pueyo, Laurent},
	editor = {Tyson, Robert K. and Lloyd-Hart, Michael},
	urldate = {2021-07-09},
	date = {2007-09-13},
	langid = {english},
}

@article{haffert_implicit_2023,
	title = {Implicit electric field conjugation: Data-driven focal plane control},
	volume = {673},
	rights = {© The Authors 2023},
	issn = {0004-6361, 1432-0746},
	url = {https://www.aanda.org/articles/aa/abs/2023/05/aa44960-22/aa44960-22.html},
	doi = {10.1051/0004-6361/202244960},
	shorttitle = {Implicit electric field conjugation},
	pages = {A28},
	journaltitle = {Astronomy \& Astrophysics},
	shortjournal = {A\&A},
	publisher = {{EDP} Sciences},
	author = {Haffert, S. Y. and Males, J. R. and Ahn, K. and Van Gorkom, K. and Guyon, O. and Close, L. M. and Long, J. D. and Hedglen, A. D. and Schatz, L. and Kautz, M. and Lumbres, J. and Rodack, A. and Knight, J. M. and Miller, K.},
	urldate = {2026-07-04},
	date = {2023-05-01},
	langid = {english},
}

@INPROCEEDINGS{poppy,
       author = {{Perrin}, Marshall D. and {Soummer}, R{\'e}mi and {Elliott}, Erin M. and {Lallo}, Matthew D. and {Sivaramakrishnan}, Anand},
        title = "{Simulating point spread functions for the James Webb Space Telescope with WebbPSF}",
    booktitle = {Space Telescopes and Instrumentation 2012: Optical, Infrared, and Millimeter Wave},
         year = 2012,
       editor = {{Clampin}, Mark C. and {Fazio}, Giovanni G. and {MacEwen}, Howard A. and {Oschmann}, Jr., Jacobus M.},
       series = {Society of Photo-Optical Instrumentation Engineers (SPIE) Conference Series},
       volume = {8442},
        month = sep,
          eid = {84423D},
        pages = {84423D},
          doi = {10.1117/12.925230},
       adsurl = {https://ui.adsabs.harvard.edu/abs/2012SPIE.8442E..3DP}
}

@inproceedings{ashcraft_2022,
	author = {Jaren N. Ashcraft and Heejoo Choi and Ewan S. Douglas and Kevin Derby and Kyle Van Gorkom and Daewook Kim and Ramya Anche and Alex Carter and Olivier Durney and Sebastiaan Haffert and Lori Harrison and Maggie Kautz and Jennifer Lumbres and Jared R. Males and Kian Milani and Oscar M. Montoya and George A. Smith},
	booktitle = {Space Telescopes and Instrumentation 2022: Optical, Infrared, and Millimeter Wave},
	doi = {10.1117/12.2628855},
	editor = {Laura E. Coyle and Shuji Matsuura and Marshall D. Perrin},
	organization = {International Society for Optics and Photonics},
	pages = {121805L},
	publisher = {SPIE},
	title = {{The space coronagraph optical bench (SCoOB): 1. Design and assembly of a vacuum-compatible coronagraph testbed for spaceborne high-contrast imaging technology}},
	url = {https://doi.org/10.1117/12.2628855},
	volume = {12180},
	year = {2022}
}

@inproceedings{maier_2020,
	author = {Erin R. Maier and Ewan S. Douglas and Dae Wook Kim and Kate Su and Jaren N. Ashcraft and James B. Breckinridge and Heejoo Choi and Elodie Choquet and Thomas E. Connors and Olivier Durney and Kerry L. Gonzales and Charlotte E. Guthery and Christian A. Haughwout and James C. Heath and Justin Hyatt and Jennifer Lumbres and Jared R. Males and Elisabeth C. Matthews and Kian Milani and Oscar M. Montoya and Mamadou N'Diaye and Jamison Noenickx and Leonid Pogorelyuk and Garreth Ruane and Glenn Schneider and George A. Smith and Christopher C. Stark},
	booktitle = {Space Telescopes and Instrumentation 2020: Optical, Infrared, and Millimeter Wave},
	doi = {10.1117/12.2560878},
	editor = {Makenzie Lystrup and Marshall D. Perrin and Natalie Batalha and Nicholas Siegler and Edward C. Tong},
	organization = {International Society for Optics and Photonics},
	pages = {324 -- 340},
	publisher = {SPIE},
	title = {{Design of the vacuum high contrast imaging testbed for CDEEP, the Coronagraphic Debris and Exoplanet Exploring Pioneer}},
	url = {https://doi.org/10.1117/12.2560878},
	volume = {11443},
	year = {2020},
}

@inproceedings{vangorkom_2024,
	title = {The space coronagraph optical bench ({SCoOB}): 4. Vacuum performance of a high contrast imaging testbed},
	volume = {13092},
	url = {https://www.spiedigitallibrary.org/conference-proceedings-of-spie/13092/1309222/The-space-coronagraph-optical-bench-SCoOB--4-Vacuum-performance/10.1117/12.3020654},
	doi = {10.1117/12.3020654},
	shorttitle = {The space coronagraph optical bench ({SCoOB})},
	eventtitle = {Space Telescopes and Instrumentation 2024: Optical, Infrared, and Millimeter Wave},
	pages = {680},
	booktitle = {Space Telescopes and Instrumentation 2024: Optical, Infrared, and Millimeter Wave},
	publisher = {{SPIE}},
	author = {Van Gorkom, Kyle and Douglas, Ewan S. and Milani, Kian and Ashcraft, Jaren N. and Anche, Ramya M. and Jenkins, Emory and Ingraham, Patrick and Haffert, Sebastiaan and Kim, Daewook and Choi, Heejoo and Durney, Olivier},
	urldate = {2026-07-04},
	date = {2024-08-23},
	langid = {english},
}

@inproceedings{kvg_2026,
	title = {The Space Coronagraph Optical Bench (SCoOB): 11.
Modeling and correction of chromatic aberrations},
	volume = {14145},
	publisher = {{SPIE}},
	author = {Van Gorkom, Kyle and others},
}

@ARTICLE{2007Natur.446..771T,
       author = {Trauger, John T. and Traub, Wesley A.},
        title = {A laboratory demonstration of the capability to image an Earth-like extrasolar planet},
      journal = {Nature},
         year = 2007,
        month = apr,
       volume = {446},
       number = {7137},
        pages = {771-773},
          doi = {10.1038/nature05729},
       adsurl = {https://ui.adsabs.harvard.edu/abs/2007Natur.446..771T}
}

@INPROCEEDINGS{2020SPIE11443E..1UK,
       author = {{Kasdin}, N. Jeremy and {Bailey}, Vanessa P. and {Mennesson}, Bertrand and {Zellem}, Robert T. and {Ygouf}, Marie and {Rhodes}, Jason and {Luchik}, Thomas and {Zhao}, Feng and {Riggs}, A.~J. Eldorado and {Seo}, Byoung-Joon and {Krist}, John and {Kern}, Brian and {Tang}, Hong and {Nemati}, Bijan and {Groff}, Tyler D. and {Zimmerman}, Neil and {Macintosh}, Bruce and {Turnbull}, Margaret and {Debes}, John and {Douglas}, Ewan S. and {Lupu}, Roxana E.},
        title = "{The Nancy Grace Roman Space Telescope Coronagraph Instrument (CGI) technology demonstration}",
    booktitle = {Space Telescopes and Instrumentation 2020: Optical, Infrared, and Millimeter Wave},
         year = 2020,
       editor = {{Lystrup}, Makenzie and {Perrin}, Marshall D.},
       series = {Society of Photo-Optical Instrumentation Engineers (SPIE) Conference Series},
       volume = {11443},
        month = dec,
          eid = {114431U},
        pages = {114431U},
          doi = {10.1117/12.2562997},
archivePrefix = {arXiv},
       eprint = {2103.01980},
 primaryClass = {astro-ph.IM},
       adsurl = {https://ui.adsabs.harvard.edu/abs/2020SPIE11443E..1UK}
}

@ARTICLE{2017JATIS...3d9002M,
       author = {{Miller}, Kelsey and {Guyon}, Olivier and {Males}, Jared},
        title = "{Spatial linear dark field control: stabilizing deep contrast for exoplanet imaging using bright speckles}",
      journal = {Journal of Astronomical Telescopes, Instruments, and Systems},
         year = 2017,
        month = oct,
       volume = {3},
          eid = {049002},
        pages = {049002},
          doi = {10.1117/1.JATIS.3.4.049002},
archivePrefix = {arXiv},
       eprint = {1703.04259},
 primaryClass = {astro-ph.IM},
       adsurl = {https://ui.adsabs.harvard.edu/abs/2017JATIS...3d9002M}
}

@ARTICLE{2021A&A...646A.145M,
       author = {{Miller}, K.~L. and {Bos}, S.~P. and {Lozi}, J. and {Guyon}, O. and {Doelman}, D.~S. and {Vievard}, S. and {Sahoo}, A. and {Deo}, V. and {Jovanovic}, N. and {Martinache}, F. and {Snik}, F. and {Currie}, T.},
        title = "{Spatial linear dark field control on Subaru/SCExAO. Maintaining high contrast with a vAPP coronagraph}",
      journal = {A\&A},
         year = 2021,
        month = feb,
       volume = {646},
          eid = {A145},
        pages = {A145},
          doi = {10.1051/0004-6361/202039583},
       adsurl = {https://ui.adsabs.harvard.edu/abs/2021A&A...646A.145M}
}

@ARTICLE{2021A&A...653A..42B,
       author = {{Bos}, S.~P. and {Miller}, K.~L. and {Lozi}, J. and {Guyon}, O. and {Doelman}, D.~S. and {Vievard}, S. and {Sahoo}, A. and {Deo}, V. and {Jovanovic}, N. and {Martinache}, F. and {Currie}, T. and {Snik}, F.},
        title = "{First on-sky demonstration of spatial Linear Dark Field Control with the vector-Apodizing Phase Plate at Subaru/SCExAO}",
      journal = {A\&A},
         year = 2021,
        month = sep,
       volume = {653},
          eid = {A42},
        pages = {A42},
          doi = {10.1051/0004-6361/202040134},
archivePrefix = {arXiv},
       eprint = {2106.06286},
 primaryClass = {astro-ph.IM},
       adsurl = {https://ui.adsabs.harvard.edu/abs/2021A&A...653A..42B}
}

@ARTICLE{2023arXiv230917449P,
       author = {{Poon}, Phillip K. and {Potier}, Axel and {Ruane}, Garreth and {Walter}, Alex B. and {Eldorado Riggs}, A J and {Noyes}, Matthew and {Mejia Prada}, Camilo and {Ahn}, Kyohoon and {Guyon}, Olivier},
        title = "{Experimental demonstration of spectral linear dark field control at NASA's high contrast imaging testbeds}",
      journal = {arXiv e-prints},
         year = 2023,
        month = sep,
          eid = {arXiv:2309.17449},
        pages = {arXiv:2309.17449},
          doi = {10.48550/arXiv.2309.17449},
archivePrefix = {arXiv},
       eprint = {2309.17449},
 primaryClass = {astro-ph.IM},
       adsurl = {https://ui.adsabs.harvard.edu/abs/2023arXiv230917449P}
}

@ARTICLE{2022JATIS...8c5001R,
       author = {{Redmond}, Susan F. and {Pogorelyuk}, Leonid and {Pueyo}, Laurent and {Por}, Emiel and {Noss}, James and {Will}, Scott D. and {Laginja}, Iva and {Brooks}, Keira and {Maclay}, Matthew and {Fowler}, J. and {Jeremy Kasdin}, N. and {Perrin}, Marshall D. and {Soummer}, R{\'e}mi},
        title = "{Implementation of a dark zone maintenance algorithm for speckle drift correction in a high contrast space coronagraph}",
      journal = {Journal of Astronomical Telescopes, Instruments, and Systems},
         year = 2022,
        month = jul,
       volume = {8},
          eid = {035001},
        pages = {035001},
          doi = {10.1117/1.JATIS.8.3.035001},
       adsurl = {https://ui.adsabs.harvard.edu/abs/2022JATIS...8c5001R}
}

@article{10.1117/1.JATIS.12.4.041014,
author = {Saikrishna Manojkumar and Susan F. Redmond and Leonid Pogorelyuk and Christine L. Page and Garreth Ruane and A. J. Eldorado Riggs and Kerri Cahoy},
title = {{Demonstration of a modal dark zone maintenance algorithm for high-contrast direct imaging}},
volume = {12},
journal = {Journal of Astronomical Telescopes, Instruments, and Systems},
number = {4},
publisher = {SPIE},
pages = {041014},
year = {2026},
doi = {10.1117/1.JATIS.12.4.041014},
URL = {https://doi.org/10.1117/1.JATIS.12.4.041014}
}

@ARTICLE{2019ApJ...873...95P,
       author = {{Pogorelyuk}, Leonid and {Kasdin}, N. Jeremy},
        title = "{Dark Hole Maintenance and A Posteriori Intensity Estimation in the Presence of Speckle Drift in a High-contrast Space Coronagraph}",
      journal = {ApJ},
         year = 2019,
        month = mar,
       volume = {873},
       number = {1},
          eid = {95},
        pages = {95},
          doi = {10.3847/1538-4357/ab0461},
archivePrefix = {arXiv},
       eprint = {1902.01880},
 primaryClass = {astro-ph.IM},
       adsurl = {https://ui.adsabs.harvard.edu/abs/2019ApJ...873...95P}
}

@inproceedings{ashcraft2024space,
  title={The space coronagraph optical bench (SCoOB): 3. Mueller matrix polarimetry of a coronagraphic exit pupil},
  author={Ashcraft, Jaren N and Douglas, Ewan S and Anche, Ramya M and {Van Gorkom}, Kyle and Jenkins, Emory and Melby, William and Millar-Blanchaer, Maxwell A},
  booktitle={Space Telescopes and Instrumentation 2024: Optical, Infrared, and Millimeter Wave},
  volume={13092},
  pages={2174--2189},
  year={2024},
  organization={SPIE}
}

@inproceedings{anche2024space,
  title={The space coronagraph optical bench (SCoOB): 5. End-to-end simulation of polarization aberrations},
  author={Anche, Ramya M and {Van Gorkom}, Kyle J and Ashcraft, Jaren N and Douglas, Ewan and Jenkins, Emory L and Haffert, Sebastiaan Y and Millar-Blanchaer, Maxwell A},
  booktitle={Space Telescopes and Instrumentation 2024: Optical, Infrared, and Millimeter Wave},
  volume={13092},
  pages={2203--2213},
  year={2024},
  organization={SPIE}
}

@inproceedings{milani2025space,
  title={The Space Coronagraph Optical Bench (SCoOB): 6. demonstration of Lyot low order wavefront control combined with high order wavefront control using a vortex coronagraph},
  author={Milani, Kian and {Van Gorkom}, Kyle and Mendillo, Christopher B and Anche, Ramya and Ashcraft, Jaren N and Derby, Kevin and Males, Jared R and Schilperoort, Adam and Douglas, Ewan S},
  booktitle={Techniques and Instrumentation for Detection of Exoplanets XII},
  volume={13627},
  pages={623--643},
  year={2025},
  organization={SPIE}
}

@inproceedings{derby2025space,
  title={The Space Coronagraph Optical Bench (SCoOB): 7. design, fabrication, and first light for a self-coherent camera},
  author={Derby, Kevin and Milani, Kian and Hathaway, Grace C and Liberman, Joshua and {Van Gorkom}, Kyle and Anche, Ramya and Schilperoort, Adam and Fucetola, Corey and Chalifoux, Brandon and Hewawasam, Kuravi and others},
  booktitle={Techniques and Instrumentation for Detection of Exoplanets XII},
  volume={13627},
  pages={644--651},
  year={2025},
  organization={SPIE}
}

@inproceedings{anche_2025,
	title = {The Space Coronagraph Optical Bench ({SCoOB}): 8. end-to-end numerical modeling of the testbed to estimate the contrast limits},
	volume = {13627},
	url = {https://www.spiedigitallibrary.org/conference-proceedings-of-spie/13627/1362726/The-Space-Coronagraph-Optical-Bench-SCoOB--8-end-to/10.1117/12.3065695},
	doi = {10.1117/12.3065695},
	shorttitle = {The Space Coronagraph Optical Bench ({SCoOB})},
	eventtitle = {Techniques and Instrumentation for Detection of Exoplanets {XII}},
	pages = {652},
	booktitle = {Techniques and Instrumentation for Detection of Exoplanets {XII}},
	publisher = {{SPIE}},
	author = {Anche, Ramya M. and {Van Gorkom}, Kyle and Milani, Kian and Derby, Kevin and Jenkins, Emory and Ashcraft, Jaren and Subramanian, Saraswathi Kalyani and Ingraham, Patrick and Kim, Daewook and Choi, Heejoo and Durney, Oli and Douglas, Ewan},
	urldate = {2026-07-04},
	date = {2025-09-18},
	langid = {english},
}

@ARTICLE{2025JATIS..11c9001M,
       author = {{Milani}, Kian and {Will}, Scott D. and {Van Gorkom}, Kyle and {Douglas}, Ewan S. and {Ashcraft}, Jaren N. and {Cahoy}, Kerri},
        title = "{Demonstrations of adjoint electric field conjugation for a vortex coronagraph}",
      journal = {Journal of Astronomical Telescopes, Instruments, and Systems},
         year = 2025,
        month = jul,
       volume = {11},
          eid = {039001},
        pages = {039001},
          doi = {10.1117/1.JATIS.11.3.039001},
       adsurl = {https://ui.adsabs.harvard.edu/abs/2025JATIS..11c9001M}
}
\bibliographystyle{spiebib} % makes bibtex use spiebib.bst

\end{document}